\documentclass[12pt]{article}
\usepackage{amsmath}
\usepackage{amssymb}
\usepackage[english]{babel}
\usepackage{graphicx}
\usepackage{subfigure}
\usepackage{indentfirst}
\usepackage{hep}
\usepackage{mathrsfs}
\usepackage{dsfont}
\usepackage{setspace}
\usepackage[labelfont=bf,labelsep=period]{caption}
\usepackage{float}
\usepackage[numbers,sort&compress]{natbib}

\begin{document}\baselineskip24pt

\title{Experimental test of the majorization uncertainty relation with mixed states}

\author
{\normalsize Shuang Wang,$^{1}$ Fang-Xia Meng$^{2}$, Hui Wang$^{1}$, and Cong-Feng Qiao$^{1,3}\footnote{E-mail: qiaocf@ucas.ac.cn}$ \\ [0.2cm]
\small{$^{1}$School of Physical Sciences, University of Chinese Academy of Sciences} \\
\small{YuQuan Road 19A, Beijing 100049, China}\\
\small{$^{2}$Center of Materials Science and Optoelectronics Engineering \& CMSOT,} \\
\small{University of Chinese Academy of Sciences, YuQuan Road 19A, Beijing 100049, China}\\
\small{$^{3}$Key Laboratory of Vacuum Physics, University of Chinese Academy of Sciences} \\
\small{YuQuan Road 19A, Beijing 100049, China}\\
}

\date{}

\maketitle

\begin{abstract}\doublespacing
The uncertainty relation lies at the heart of quantum theory and behaves as a non-classical constraint on the indeterminacies of  incompatible observables in a system. In the literature, many experiments have been devoted to the test of the uncertainty relations which mainly focus on the pure states. In this work we test the novel majorization uncertainty relations of three incompatible observables using a series of mixed states with adjustable mixing degrees, and compare the compactness of various entropy uncertainty relations. The experimental results confirm that for general mixed quantum system, the majorization uncertainty relation tends to be  the tightest constraint on uncertainty, and indicate that the entropy uncertainty relation obtained from the majorzation uncertainty relation is the optimal one. Our experimental setup provides an easy means for preparing mixed states, and based on this simple optical elements can be utilized to realize the required quantum states.
\end{abstract}

\newpage

\section{Introduction}

The concept of uncertainty principle was initially proposed by Heisenberg through a gedanken experiment, where the canonically conjugate quantities, $x$ and $p$, are shown to be determined simultaneously only with a characteristic indeterminacy \cite{heis}. One of the well-known formulations of the uncertainty relation reads \cite{Robertson}
\begin{equation}
\Delta X^2 \Delta Y^2 \geqslant
\frac{1}{4}| \langle [X,Y] \rangle |^2 \; .   \label{1}
\end{equation}
Here the variance $\Delta X^2 = \langle X^2\rangle -\langle X\rangle^2$ measures the uncertainty of the observable and $[X,Y]\equiv XY-YX$ is the commutator. Another measure of the uncertainty, the entropy, was also explored in the study of uncertainty relations \cite{Entropic-Un-1} with a typical form \cite{Entropic-Un-2}
\begin{equation}
H(\vec{p}\,) +H(\vec{q}\,) \geqslant \log_{2} \frac{1}{c} \; , \label{2}
\end{equation}
where $H(\vec{p}\,) \equiv -\sum_{i} p_{i}\log_{2} p_{i}$ is the Shannon entropy for the probability distribution of measuring $X$ with the outcomes $x_i$ ($\vec{q}$ is similarly defined for $Y$), and $c \equiv \max_{i,j}|\langle x_i|y_j\rangle|^2$ is the maximum overlap between the eigenvectors of $X$ and $Y$. Recently, the ``universal uncertainty relation'' in terms of majorization relation was proposed \cite{Maj-1,Maj-2}, and an optimal majorization uncertainty relation in direct sum, i.e.
\begin{equation}
\vec{p} \oplus \vec{q} \prec \vec{s} \; , \label{3}
\end{equation}
was obtained \cite{Li-Optimal}. Here the vector $\vec{s}$ is constructed from the lattice theory and depends only on the observables $X$ and $Y$. The majorization relation $\vec{p} \prec \vec{q}$ means $\sum_{\mu=1}^k p^{\downarrow}_{\mu} \leqslant \sum_{\nu=1}^k q^{\downarrow}_{\nu}$, $\forall k\in \{1,\cdots, N\}$ with equality satisfied when $k=N$. $\oplus$ is the direct sum operation. The direct sum of an $m$-dimensional vector  $\vec{p}$  and an $n$-dimensional vector  $\vec{q}$  is an $(m+n)$-dimensional vector obtained by arranging the components of  $\vec{p}$  and  $\vec{q}$  into a single column vector. The superscript $\downarrow$ indicates that the components are arranged in descending order.

Practically, the states mostly encountered are generally mixed. However, tightening the bounds for mixed states turns out to be difficult for both variance and entropic uncertainty relations. The variance-based uncertainty relation characterizing the full range of uncertainty regions for mixed states was only explicitly constructed for qubit system \cite{Li-Reformulating}. The quantum-memory-assisted uncertainty relation leads to an improvement of relation (\ref{2}) \cite{Memory-Uncertianty1, Memory-Uncertainty2}
\begin{equation}
H(\vec{p}\,) + H(\vec{q}\,) \geqslant \log_2 \frac{1}{c}+ S(\rho) \; , \label{4}
\end{equation}
where $S(\rho) \equiv -\mathrm{Tr}[\rho\log_{2}\rho] = H( \vec{\lambda} \,)$ is the von Neumann entropy with $\vec{\lambda}$ being the vector composed of the eigenvalues of density matrix $\rho$ of quantum state. Nevertheless, to get tighter bounds for high-dimensional multi-observables is still a challenging task in the study of the entropic uncertainty relation \cite{Entropy-RMP, Entropy-2019, Entropy-3-mem}. On the contrary, the optimal bound problem for direct sum majorization uncertainty relations had been solved for general mixed states by exploring the lattice theory \cite{Li-Optimal}. For example, in qubit systems there exists the following relation \cite{Li-Optimal}
\begin{equation}
\bigoplus_{i=1}^{N} \vec{p}_{M_{i}} \prec \left(\lambda_1\vec{s}^{\,\uparrow} + \lambda_2\vec{s}^{\,\downarrow}\right) \; . \label{5}
\end{equation}
Here, $M_i$ are $N$-measurement operators and $\lambda_1\leqslant \lambda_2$ are the eigenvalues of $\rho$. The superscript $\uparrow$ means the components being in ascending order.

Existing tests of the uncertainty relations mainly focus on the pure systems. The variance-based uncertainty relations have been tested with single quanta of SPDC photons \cite{Xue-1, zhixin1, zhixin2, Xue-2, Reverse-UR} and nitrogen-vacancy (NV) center system \cite{dufei}. The entropic uncertainty relation for multiple observables was also investigated in the NV center system \cite{NV-entropy}. Refs. \cite{Single-Maj-exp1, Single-Maj-exp2} tested the majorization uncertainty relation using the SPDC photon sources. The investigation of the direct sum majorization uncertainty relation using the coherent lights was carried out in Ref. \cite{WangH} by measuring the Lorenz curves. For mixed system, a test of the variance-based and the related entropic uncertainty relations of two qubit observables were carried out using the neutron spins \cite{Neu-Mixed}. As actual systems are generally mixed, it is important to verify the various uncertainty relations within the mixed states, from which the practical effects of the uncertainty relation on precision measurements in a noisy world may also be obtained.

In this work, we present an experimental investigation of the optimal majorization uncertainty relation for three incompatible observables with varying degrees of mixedness using the coherent light. We prepare the mixed polarization states with different degrees of mixedness by combining two linearly polarized coherent beams. The optimization of the direct sum majorization uncertianty relation for general states is explored. By applying the entropic functions to the majorization relation, the lower bounds for several typical entropic uncertainty relations applicable to mixed states are compared via experiment.

\section{Uncertainty relations with Stokes parameters}

Let $\rho$ be the $2\times 2$ density matrix for the polarization degrees of the light, then the Stokes parameters, which fully characterize the polarization state of the light, are defined by \cite{Book-stokes}
\begin{align}
S_0 & \equiv \mathcal{N}( \langle H| \rho |H\rangle + \langle V| \rho |V\rangle ) \; , \;
S_1 \equiv \mathcal{N}( \langle H|\rho |H\rangle - \langle V| \rho |V\rangle ) \; ,\label{6}\\
S_2 & \equiv \mathcal{N}(\langle +|\rho |+\rangle  - \langle -| \rho |- \rangle) \; , \;
S_3 \equiv \mathcal{N}(\langle R| \rho |R\rangle -\langle L| \rho |L\rangle) \; . \label{7}
\end{align}
Here $\mathcal{N}$ is a constant depending on the detector efficiency and light intensity; $|H/V\rangle$, $|\pm\rangle = \frac{1}{\sqrt{2}}( |H\rangle \pm |V\rangle)$, and $|L/R\rangle = \frac{1}{\sqrt{2}}( |H\rangle \pm \mathrm{i}|V\rangle)$ are kets for linear horizontal/vertical, linear diagonal($\pm 45^{\circ}$), and left/right circular polarizations. In terms of Stokes parameters, the density matrix may be expressed as \cite{James}
\begin{equation}
\rho = \frac{1}{2} \sum_{i=0}^3 \frac{S_i}{S_0} \hat{\sigma}_i  \; , \label{8}
\end{equation}
where $\hat{\sigma}_{0}= I$, $\hat{\sigma}_1 = |H\rangle\langle H|- |V\rangle \langle V|$, $\hat{\sigma}_2 = |H\rangle\langle V| + |V\rangle \langle H|$, and $\hat{\sigma}_3 = i|H\rangle\langle V|- i|V\rangle \langle H|$. By identifying $\sigma_z = \hat{\sigma}_1$, $\sigma_y=-\hat{\sigma}_3$, and $\sigma_x=\hat{\sigma}_2$, Pauli matrices can be formulated in $|H/V\rangle$ base, with expectation values $\langle \sigma_x\rangle = S_2/S_0$, $\langle \sigma_y\rangle = -S_3/S_0$, and $\langle \sigma_z\rangle = S_1/S_0$.

In two-dimensional systems, the expectation value of a dichotomic observable can be expressed as $\langle \sigma_i \rangle = p_{i+} - p_{i-} $, $i=x,y,z$, where $\pm $ denote the two outcomes of $\sigma_i$. Taking $\sigma_z$ as an example, we have
\begin{align}
p_{z+} = \frac{1+\langle \sigma_z\rangle}{2} = \frac{S_0+S_1}{2S_0}\; , \; p_{z-} = \frac{1- \langle \sigma_z\rangle}{2} = \frac{S_0-S_1}{2S_0} \; .\label{10}
\end{align}
Similar calculations apply to $\sigma_x$ and $\sigma_y$ as well.

For two observables, we can obtain a new entropy uncertainty relation employing the majorization uncertainty relation (\ref{5}), and compare it with the prevailing entropy uncertainty relation (\ref{4}) in the context of mixed state.

Consider two observables
\begin{equation}
Z = \left(\begin{array}{cc}
{1} & {0} \\
{0} & {-1}
\end{array}\right) \ {\rm and}\ X(\theta)=\left(\begin{array}{cc}
{\cos \theta} & {\sin \theta} \\
{\sin \theta} & {-\cos \theta}
\end{array}\right)\label{18}
\end{equation}
with $\theta \in (0, \frac{\pi}{2} ]$, since $Z=\sigma_{z}$ and $X (\theta)= \sin\theta\sigma_x+\cos\theta\sigma_z$, then $[Z,X(\theta)]=2i\sin\theta \sigma_y$. For a two-dimensional qubit system, when measuring an observable, e.g. $X(\theta)$, one may get values $\pm 1$ with probabilities $p_{\;_{\theta \pm}}$. Hence $\langle X(\theta) \rangle = p_{\;_{\theta+}}-p_{\;_{\theta-}}$ is the expectation value. Considering of the normalization,  we can immediately get
\begin{align}
p_{\;_{\theta+}}=\frac{1+\langle X(\theta)\rangle}{2} = \frac{S_0+S(\theta)}{2S_0} \; , \;
p_{\;_{\theta-}}=\frac{1-\langle X(\theta)\rangle}{2} =\frac{S_0-S(\theta)}{2S_0} \;.\label{19}
\end{align}
Here, $S(\theta)=\cos \theta S_{1}+\sin \theta S_{2}$.

For observables $X$ and $Z$ in (\ref{18}), uncertainty relations (\ref{6}) and (\ref{4}) are expressed as
\begin{align}
H(\vec{p}_{\theta}) +H(\vec{p}_z) & \geq \log_{2} \frac{1}{c} + H(\vec{\lambda}) \; , \label{20} \\
H(\vec{p}_{\theta}) +H(\vec{p}_z) & \geq H(\lambda_1\vec{s}^{\,\downarrow} + \lambda_2 \vec{s}^{\,\uparrow}) \; , \label{21}
\end{align}
where $\vec{\lambda} = (\lambda_1,\lambda_2)$ is eigenvalues of $\rho$, $\vec{s} = \displaystyle (1, \cos \frac{\theta}{2}, 2  \sin ^{2} \frac{\theta}{4}, 0)$.

For three observables, the direct sum majorization relation for $\sigma_x$, $\sigma_y$, and $\sigma_z$ reads \cite{Li-Optimal}
\begin{equation}
\vec{p}_{x} \oplus \vec{p}_{y} \oplus \vec{p}_{z} \prec \vec{s}(\rho)=\lambda_{1} \vec{s}^{\,\uparrow}+\lambda_{2} \vec{s}^{\,\downarrow} \; , \label{11}
\end{equation}
where $\lambda_{1,2}$ are eigenvalues of the density matrix $\rho$, $\vec{s}=\left(1, \frac{\sqrt{2}}{2}, \frac{1+\sqrt{3}- \sqrt{2}}{2}, \frac{1-\sqrt{3}+ \sqrt{2}}{2}, \frac{2-\sqrt{2}}{2}, 0\right)^{\mathrm{T}}$ and $\vec{p}_i=(p_{i+}, p_{i-})^{\mathrm{T}}$.
Expressing (\ref{11}) by Stokes parameters, we have
\begin{equation}
\vec{P}=\frac{1}{2 S_{0}}\left[\left(\begin{array}{c}
S_{0}+S_{1} \\
S_{0}-S_{1}
\end{array}\right) \oplus\left(\begin{array}{c}
S_{0}+S_{2} \\
S_{0}-S_{2}
\end{array}\right) \oplus\left(\begin{array}{c}
S_{0}-S_{3} \\
S_{0}+S_{3}
\end{array}\right)\right] \prec \lambda_{1} \vec{s}^{\,\uparrow}+\lambda_{2} \vec{s}^{\,\downarrow}\; .\label{12}
\end{equation}
Here the eigenvalues $\lambda_{1,2}$ characterize the mixedness of density matrices, e.g., $\lambda_2=1,\lambda_1=0$ when $\rho$ is pure and $\lambda_2=\lambda_1=1/2$ for completely mixed state. We define the Lorentz curve function $f(n,\vec{P})=\sum_{n}^{6}\vec{P}^{\,\downarrow}(n)$. Given two 6-dimensional vectors $\vec{r}$ and $\vec{s}$ satisfying $\vec{r}\prec\vec{s}$, we then have $f(n,\vec{r}) \leqslant f(n,\vec{s})$ with the equality hold when n=6. Meanwhile, note that the Lorentz curve $f(n,\vec{s})$ will completely wrap the Lorentz curve $f(n,\vec{r})$.

In the literature we find three types of entropic uncertainty relations about three observables defined in Stokes parameters \cite{Entropy-lmf,Entropy-rpz}, i.e.,
\begin{equation}
E \geqslant\frac{3}{2}S(\rho)-\frac{1}{2} \log _{2} \frac{1}{8}\; , \label{13}
\end{equation}
\begin{equation}
E \geqslant-\sum l_{i} \log _{2} l_{i}\; , \label{14}
\end{equation}
with $l=\left\{1, \frac{\sqrt{2}}{2}, \frac{1+\sqrt{5}-\sqrt{2}}{2}, \frac{1+\sqrt{10}-\sqrt{5}}{2}, \frac{1+\sqrt{17}-\sqrt{10}}{2}, \frac{1-\sqrt{17}}{2}\right\}$, and
\begin{equation}
E \geqslant 2\left(-\lambda_{1} \log _{2} \lambda_{1}-\lambda_{2} \log _{2} \lambda_{2}\right)-\log _{2} \frac{1}{2}\ , \label{15}
\end{equation}
where $E=-\sum_{i j} p_{j_{i}} \log_{2} p_{j_{i}}$ ($j=x, y, z$ and $ i=1,2$).
A different type of entropic uncertainty relation can be obtained from the majorization relation Eq.(\ref{11}), that is
\begin{equation}
H\left(\vec{p}_{x}\right)+H\left(\vec{p}_{y}\right)+H\left(\vec{p}_{z}\right) \geqslant H\left(\lambda_{1} \vec{s}^{\, \uparrow}+\lambda_{2} \vec{s}^{ \,\downarrow}\right)\; ,
\label{16}
\end{equation}
which in Stokes parameters reads
\begin{equation}
E \geqslant-\sum_{i} s_{i}(\lambda) \log _{2} s_{i}(\lambda)\; , \label{17}
\end{equation}
where $\vec{s}(\lambda)=\lambda_{1} \vec{s}^{\,\uparrow}+\lambda_{2} \vec{s}^{\,\downarrow}$.

\section{Experiment with coherent light}

\subsection{The state preparation and experiment setup}
\begin{figure} \centering \includegraphics[width=1.0\textwidth]{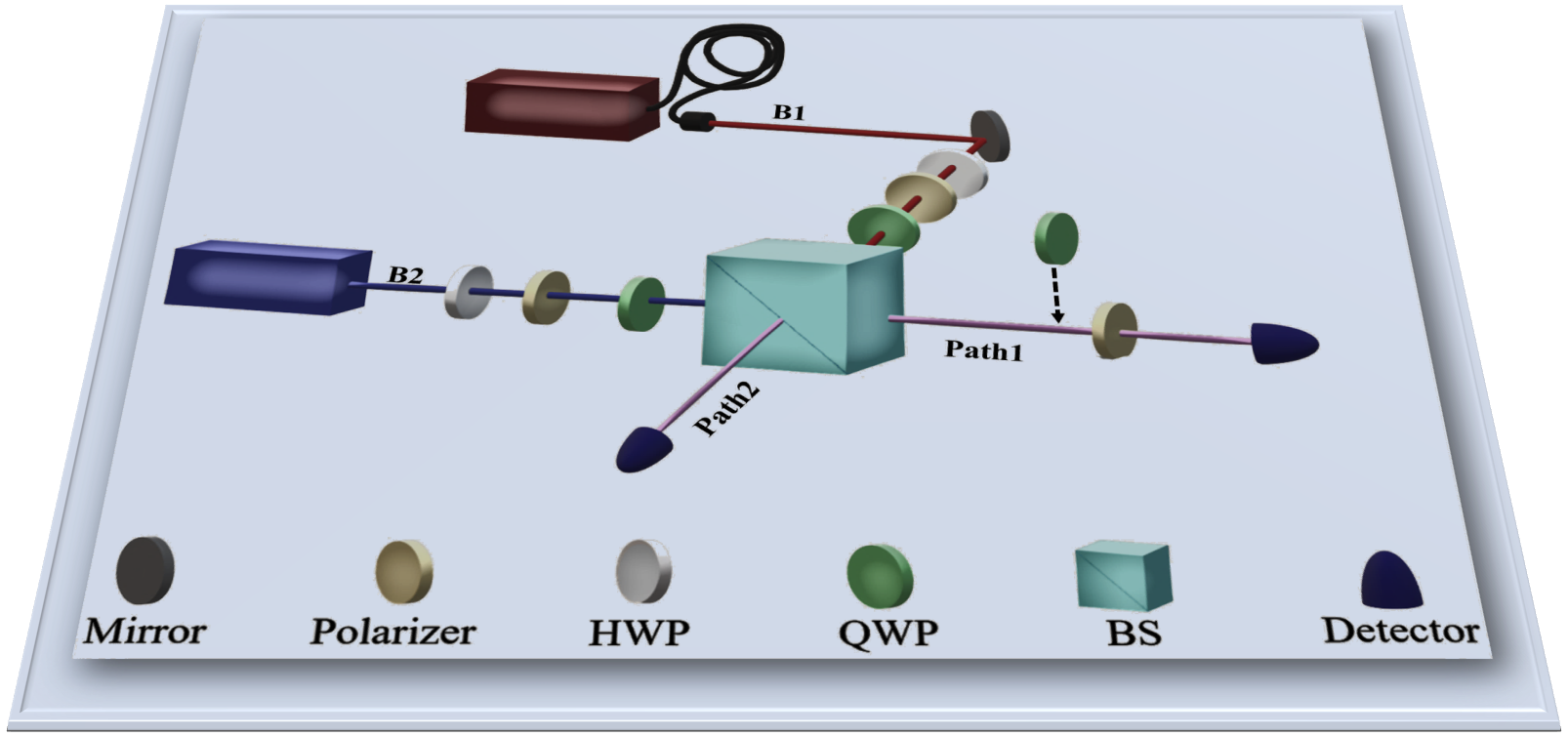}
	\caption{{\bf Experimental setup.} The two laser beams with different frequencies are respectively prepared in two mutually orthogonal polarization states through a series of devices and combined at the beam splitter. The measurement of Stokes parameters is performed in Path 1, and the light of Path 2 is detected as reference light.}\label{f1}
\end{figure}

In the experiment, polarized mixed states were prepared by means of two linearly polarized laser beams, i.e., 808 nm (B1) and 780 nm (B2), as schematically shown in Figure \ref{f1}. Before combining the two beams at the beam splitter, B1 was prepared in $|\psi\rangle = \cos \alpha|H\rangle+\sin \alpha|V\rangle$ state, while B2 was prepared in $|\psi^{\perp}\rangle$ state perpendicular to $|\psi\rangle$. Since at the detection port of path 1 (P1) $I_{P1} = I_{B1} +  I_{B2}$, when a polarimeter $|x\rangle \langle x|$ is placed in front of the detector of P1, we may get
\begin{align}
I_{P1}^{(X)} =  I_{B1}|\langle x|\psi\rangle |^2 + I_{B2}|\langle x|\psi^{\perp}\rangle |^2 \; , \label{27}
\end{align}
which was once verified \cite{ws}. It is easy to find that the probability of observing polarization $|x\rangle$ in combined beam $I_{P1}$ is
\begin{align}
p_{x} \equiv I_{P1}^{(X)}/I_{P1} = \mathrm{Tr} \left[ |x\rangle \langle x| \rho  \right] \; ,
\end{align}
where $\rho \equiv \left(I_{B1}| \psi\rangle\langle \psi| +  I_{B2}|\psi^{\perp}\rangle \langle \psi^{\perp}|\right)/I_{P1}$ is a normalized polarization state whose mixedness is determined by the intensities $I_{B1}$ and $I_{B2}$. The state $\rho$ can be re-expressed as
\begin{align}
\rho ( \lambda_{1},\alpha ) = \lambda_{1} |\psi \rangle \langle \psi |+\lambda_{2} | \psi^{\perp}\rangle\langle\psi^{\perp}|\;
\end{align}
with $\lambda_i = I_{Bi}/I_{P1}$($i$ = 1 or 2). Note, only one of the $\lambda_i$ is independent, parameter $\alpha$ appears in the prepared state$ \left| \psi\right\rangle $.

In our experiment, the detector has a response time of 1 $\mu s$ and the resolution of the Power meter is 100 pW, which are enough to identify the change of light intensity to be measured. Moreover, only the linear polarization is measured. The half-wave plate was placed before the polarizer to play the same role as an attenuator, employed to control the ratio of the light intensity of the two beams entering the beam splitter; the quarter-wave plate was used to compensate the polarization change caused by the beam splitter (BS). A series of mixed states $\rho ( \lambda_{1},\alpha )$ were prepared with $\lambda_{1}= \displaystyle \left\{ 0, 0.2, 0.3, 0.4, 0.5 \right\}$ and $\alpha= \left\{\displaystyle 0, \frac{\pi}{12}, \frac{\pi}{6}, \frac{\pi}{4}, \frac{\pi}{3}, \frac{5\pi}{12}\right\}$. By dint of $F\left(\rho, \rho_{1}\right)=\operatorname{tr}\left(\rho \rho_{1}\right)+\sqrt{1-\operatorname{Tr}\left(\rho^{2}\right)} \sqrt{1-\operatorname{Tr}\left(\rho_{1}^{2}\right)}$ \cite{fidelity}, we evaluated the fidelity of the prepared states which we found exceeded 98\%. Moreover, (\ref{27}) was verified to ensure that the two beams were collimated in producing stable mixed polarization states.

\subsection{Measuring the majorization uncertainty relations}

\begin{figure}[!hbtp]
\centering
\includegraphics[width=0.7\textwidth]{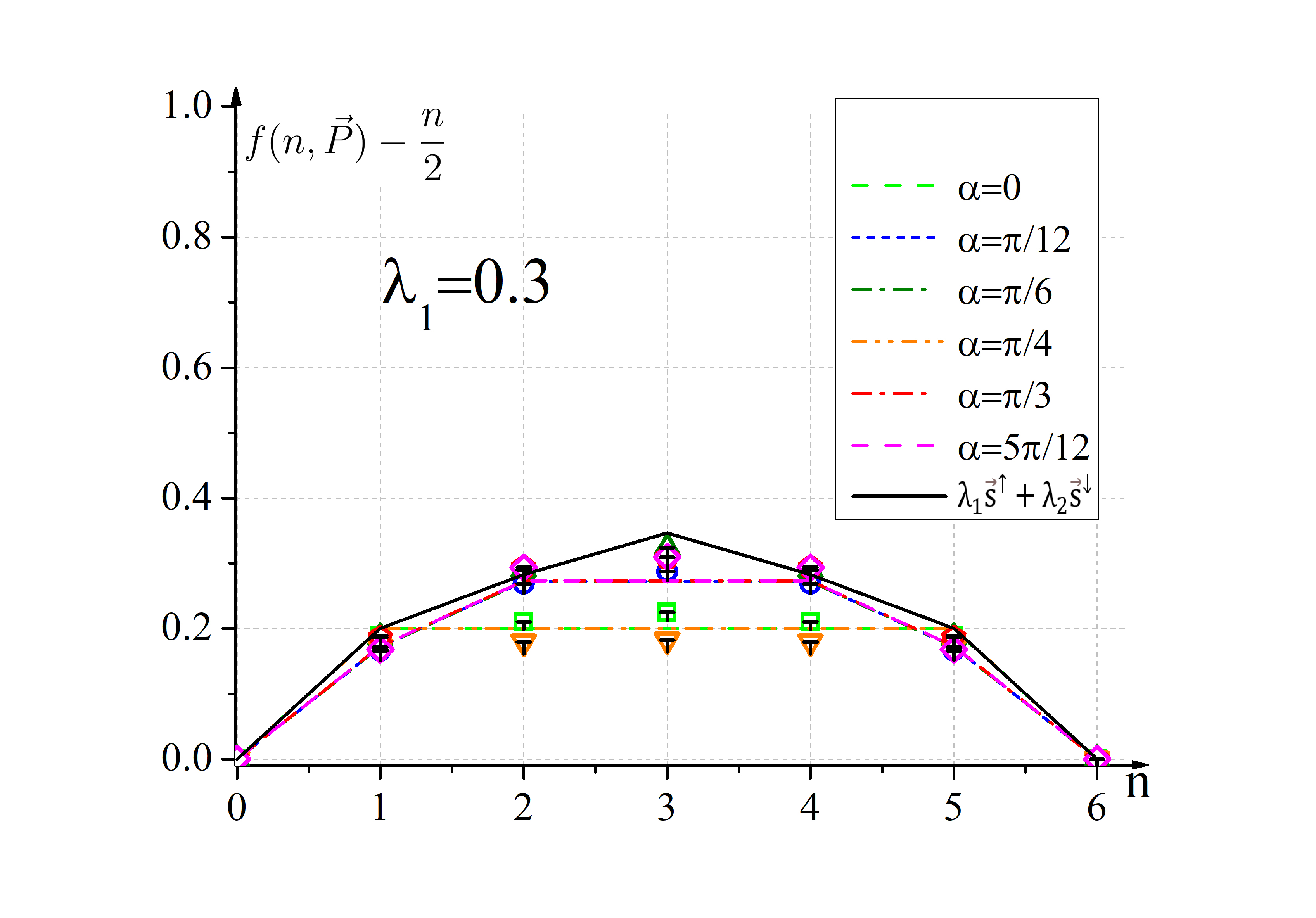}
\caption{{\bf Experimental results.} Experimental results for $\lambda_{1}=0.3$ and $\alpha=\left\{0, \frac{\pi}{12}, \frac{\pi}{6}, \frac{\pi}{4}, \frac{\pi}{3}, \frac{5 \pi}{12}\right\}$. The solid black curve in the figure corresponds to the Lorentz curve of the vector $\lambda_1\vec{s}^{\,\uparrow} + \lambda_2 \vec{s}^{\,\downarrow}$, and the dashed curves of green, blue, olive,  orange, red, and magenta correspond to the Lorentz curves of theoretical probability distributions at $\alpha= 0,\frac{\pi}{12},\frac{\pi}{6},\frac{\pi}{4},\frac{\pi}{3},\frac{5\pi}{12}$ respectively. The hollow shapes in same colors as the curves are experimental data points in corresponding states. Black short lines are error bars with $\pm 1$$\sigma$ standard deviation.}\label{f2}
\end{figure}

\begin{figure}[!hbtp] \centering
\includegraphics[width=0.85\textwidth]{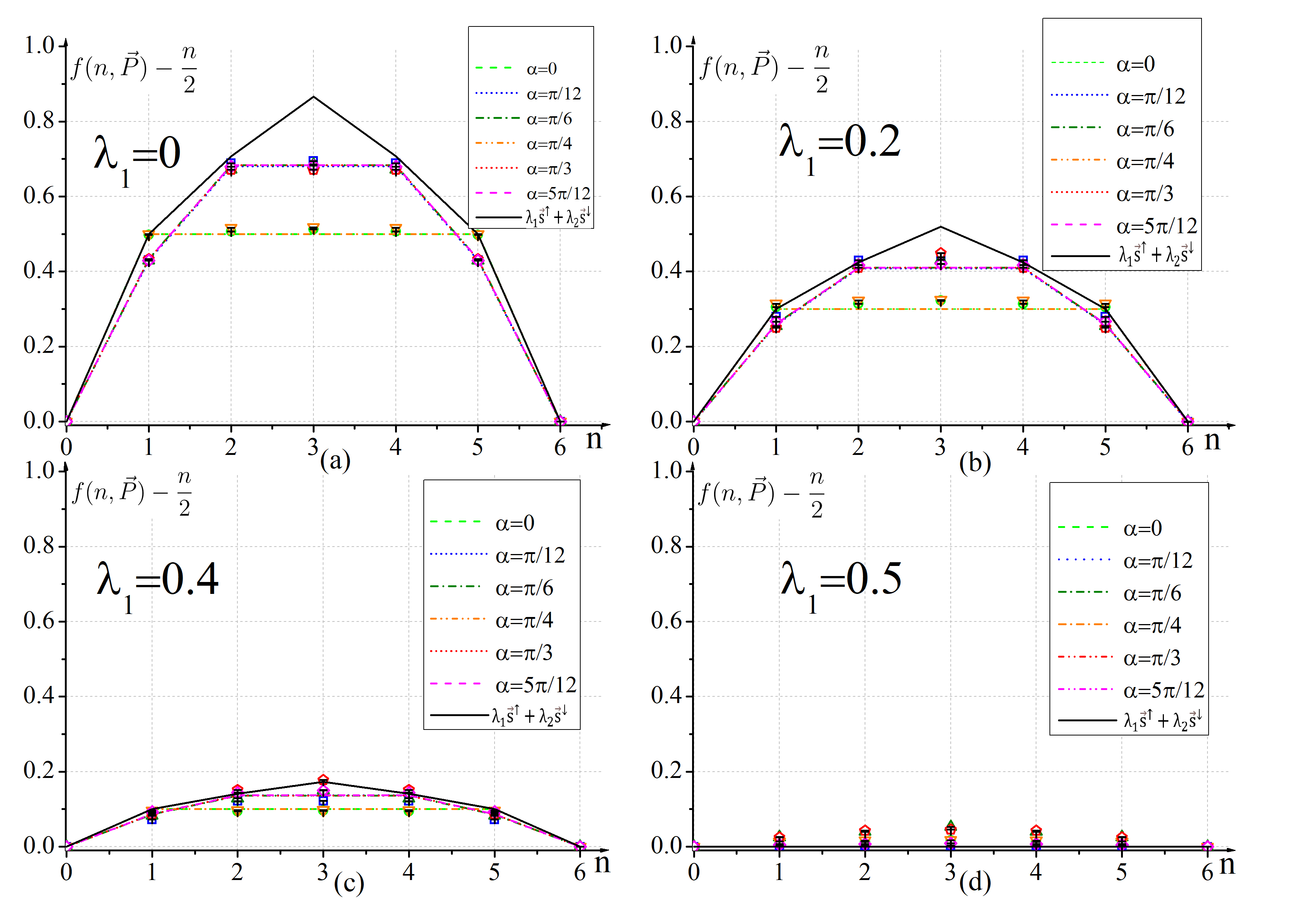}
\caption{{\bf Experimental results.} Experimental results for $\lambda_{1}= \left\{0,0.2,0.4,0.5\right\}$ and $\alpha=\left\{0, \frac{\pi}{12}, \frac{\pi}{6}, \frac{\pi}{4}, \frac{\pi}{3}, \frac{5 \pi}{12}\right\}$. The solid black curve in the figure corresponds to the Lorentz curve of the vector $\lambda_1\vec{s}^{\,\uparrow} + \lambda_2 \vec{s}^{\,\downarrow}$, and the dashed curves of green, blue, olive,  orange, red, and magenta correspond to the Lorenz curves of the theoretical probability distributions of $\alpha= 0,\frac{\pi}{12},\frac{\pi}{6},\frac{\pi}{4},\frac{\pi}{3},\frac{5\pi}{12}$ in sequence. The hollow shapes of the same colors as the curves are the experimental values in corresponding states. Black short lines are error bars with $\pm 1$$\sigma$ standard deviation.}\label{f3}
\end{figure}

We measure observables $\sigma_{x}$, $ \sigma_{y}$, and $\sigma_{z}$ under the prepared states $\rho(\lambda_{1}, \alpha)$ with $\lambda_{1} = \left\{0,0.2,0.3,0.4,0.5\right\}$ and $\alpha=\left\{0, \frac{\pi}{12}, \frac{\pi}{6}, \frac{\pi}{4}, \frac{\pi}{3}, \frac{5 \pi}{12}\right\}$ respectively.  The curves corresponding to the vector $\lambda_1\vec{s}^{\,\uparrow} + \lambda_2 \vec{s}^{\,\downarrow}$ in all figures wrap those curves corresponding to the probability distribution vectors in specific states. The experimental data points almost fall on or around the theoretical curve. The wrapping relationship between Lorentz curves of boundary vectors and other Lorentz curves in Figures 2 and 3 is the same. There are three main reasons for the drift of experimental data points relative to theoretical expected value, namely, the imperfectness of calibration , the retardation of wave plate and the fluctuation of light sources. The relation (\ref{11}) is verified with any polarized mixed state. Note, some data points in Figure \ref{f2} and Figure \ref{f3} sit in the boundary curves (the black lines), indicating that the relation (\ref{11}) is really optimal.

\subsection{Measuring the entropic functions}

\begin{figure}[!hbtp] \centering
\includegraphics[width=1\textwidth]{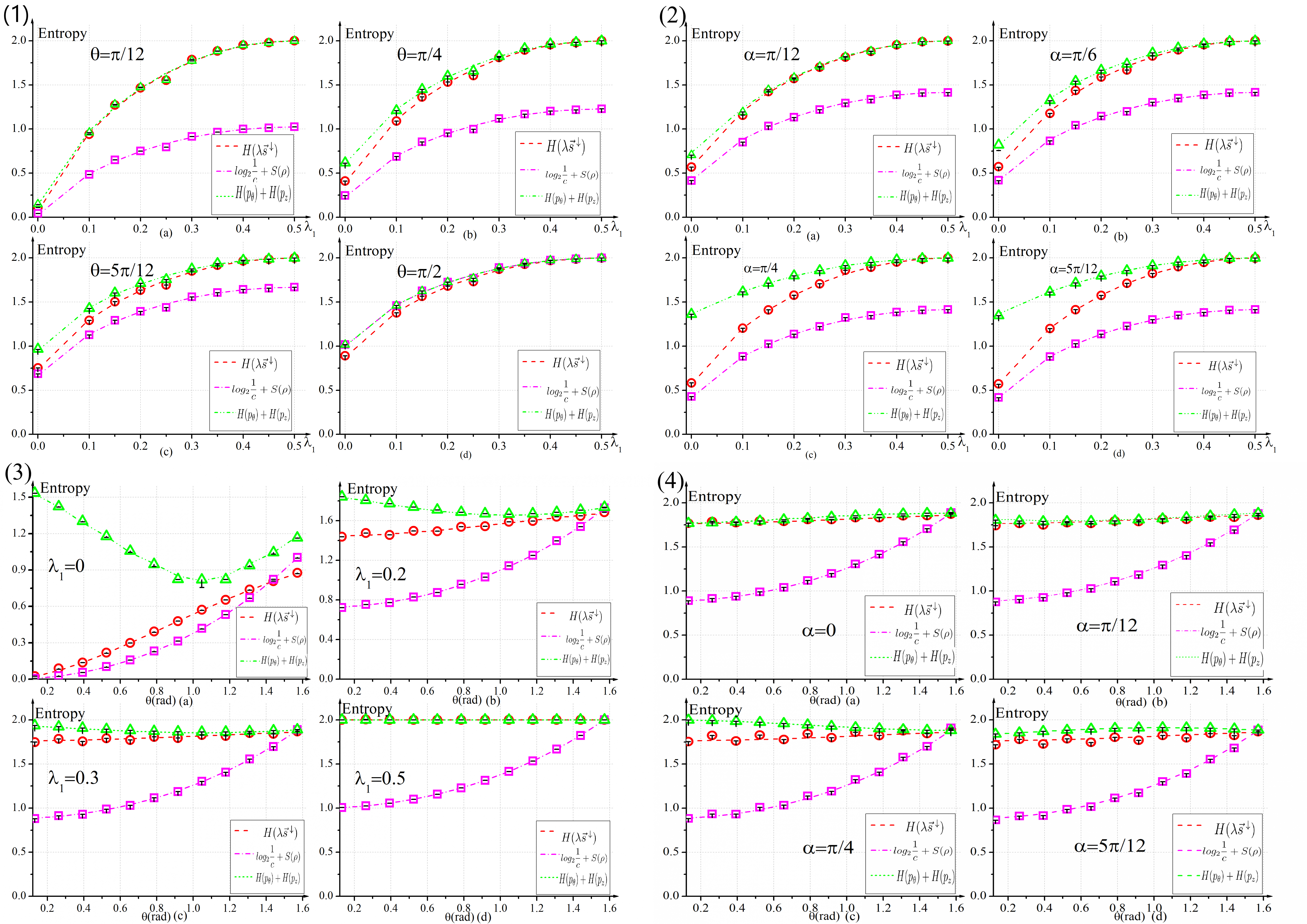}
\caption{{\bf Experimental results.} Diagram (1) shows the experimental results for $\lambda_{1} = \left\{0,0.2,0.3,0.4,0.5 \right\}$ and $\alpha=0$. (a) for $X( \frac{\pi}{12}) $ and $\sigma_{z}$, (b) for $X(\frac{\pi}{4})$ and $\sigma_{z}$, (c) for $X(\frac{5\pi}{12})$ and $\sigma_{z}$, and (d) for $X(\frac{\pi}{2})$ and $\sigma_{z}$. Diagram (2) shows the experimental results of measuring $\sigma_z$ and $X(\frac{\pi}{3})$ with various states. (a) for states $\rho(\lambda_{1},\frac{\pi}{12})(\lambda_{1}\in\{0,0.2,0.3,0.4,0.5\})$, (b) for states  $\rho(\lambda_{1},\frac{\pi}{6})(\lambda_{1}\in\{0,0.2,0.3,0.4,0.5\})$, (c) for states $\rho(\lambda_{1},\frac{\pi}{4})(\lambda_{1}\in\{0,0.2,0.3,0.4,0.5\})$, and (d) for states $\rho(\lambda_{1},\frac{5\pi}{12})(\lambda_{1}\in\{0,0.2,0.3,0.4,0.5\})$. Diagram (3) shows the experimental results for $\alpha=\frac{\pi}{6}$ and eleven pairs of observables $\{\sigma_{z},X(\theta)\}$$(\theta=\frac{\pi}{24},\frac{\pi}{12},\frac{\pi}{8},\frac{\pi}{6}, \frac{5\pi}{24},\frac{\pi}{4},\frac{7\pi}{24},\frac{\pi}{3},\frac{3\pi}{8},\frac{5\pi}{12},\frac{11\pi}{24},\frac{\pi}{2})$. (a) for state $\rho(0,\frac{\pi}{6})$, (b) for state $\rho(0.2,\frac{\pi}{6})$, (c) for state $\rho(0.3,\frac{\pi}{6})$, and (d) for state $\rho(0.5,\frac{\pi}{6})$. Diagram (4) shows the experimental results for $\lambda_{1}=0.3$ and eleven pairs of observables $\{\sigma_{z},X(\theta)\}$$(\theta=\frac{\pi}{24},\frac{\pi}{12},\frac{\pi}{8},\frac{\pi}{6}, \frac{5\pi}{24},\frac{\pi}{4},\frac{7\pi}{24},\frac{\pi}{3},\frac{3\pi}{8},\frac{5\pi}{12},\frac{11\pi}{24},\frac{\pi}{2})$. (a) for state $\rho(0.3,0)$, (b) for state $\rho(0.3,\frac{\pi}{12})$, (c) for state $\rho(0.3,\frac{\pi}{4})$, and (d) for state $\rho(0.3,\frac{5\pi}{12})$. Magenta dashed curve and red dashed curve represent the right sides of inequalities (\ref{20}) and (\ref{21}), respectively. The dotted green curves signify the theoretical calculations of $H\left(\vec{p}_{\theta}\right)+H\left(\vec{p}_{z}\right)$ respectively for states $\rho(\lambda_{1},0)(\lambda_{1}\in\{0,0.2,0.3,0.4,0.5\})$. The hollow shapes of the same colors as the curves are the experimental data points in corresponding states. The error bars indicate $\pm 1$$\sigma$ standard deviation.
}\label{f4}
\end{figure}


Diagrams \ref{f4}(1) and \ref{f4}(2) show the experimental results of entropy uncertainty relations when the mixing degree of the mixed state system changes regularly. Under the mixed-state systems $\rho\left(\lambda_{1},0 \right) $ ($\lambda_{1}\in\left\lbrace0,0.2,0.3,0.4,0.5 \right\rbrace $), four pairs of observables ($\sigma_{z}$ and $X\left( \frac{\pi}{12}\right)$, $\sigma_{z}$ and $X\left( \frac{\pi}{4}\right)$, $\sigma_{z}$ and $X\left( \frac{5\pi}{12}\right)$, and $\sigma_{z}$ and $X\left( \frac{\pi}{2}\right)$) are measured, and the results are shown in Diagram \ref{f4}(1). One may notice that the experimental data generally agree with theoretical predictions, which indicates that the two entropy uncertainty relations (\ref{20}) and (\ref{21}) both work well. Obviously, the entropy uncertainty relation (\ref{21}) is superior to (\ref{20}) in most cases investigated. But for the observables $\sigma_{z}$ and $X\left( \frac{\pi}{2}\right)$, the constraint of relation (\ref{20}) is a bit better (tighter) than (\ref{21}). For further investigation, we measure observables $\sigma_{z}$ and $X\left( \frac{\pi}{3}\right)$ under the circumstances of various mixed states, i.e. $\rho\left(\lambda_{1},\frac{\pi}{12} \right) $, $\rho\left(\lambda_{1},\frac{\pi}{6} \right) $, $\rho\left(\lambda_{1},\frac{\pi}{4} \right) $ and $\rho\left(\lambda_{1},\frac{5\pi}{12} \right) $, with $\lambda_{1} = \left\{0,0.2,0.3,0.4,0.5\right\}$. The results are shown in diagram \ref{f4}(2). Note as well, in these cases the entropy uncertainty relations (\ref{20}) and (\ref{21}) still hold, and the relation (\ref{21}) tends to be tighter than (\ref{20}).

Diagrams \ref{f4}(3) and \ref{f4}(4) exhibit the experimental results in measuring the entropy uncertainty relations on different observables. In diagram \ref{f4}(3), we prepare four mixed states $\rho\left(0,0 \right)$, $\rho\left(0,\frac{\pi}{12} \right)$, $\rho\left(0,\frac{\pi}{4} \right)$ and $\rho\left(0,\frac{5 \pi}{12} \right)$. The experimental data are in good agreement with the theoretical predictions, which indicate again the validity of the entropy uncertain relations (\ref{20}) and (\ref{21}). It is obvious that the latter is more stringent than the former in these states. Diagram \ref{f4}(4) measures the same stuff as Figure \ref{f4}(3) but with another four mixed states $\rho\left(0,\frac{\pi}{6}\right)$, $\rho\left(0.2,\frac{\pi}{6} \right)$, $\rho\left(0.3,\frac{\pi}{6} \right)$ and $\rho\left(0.2,\frac{\pi}{6} \right)$, and similar conclusions are obtained.

\begin{figure}[!hbtp]
\centering
\includegraphics[width=0.9\textwidth]{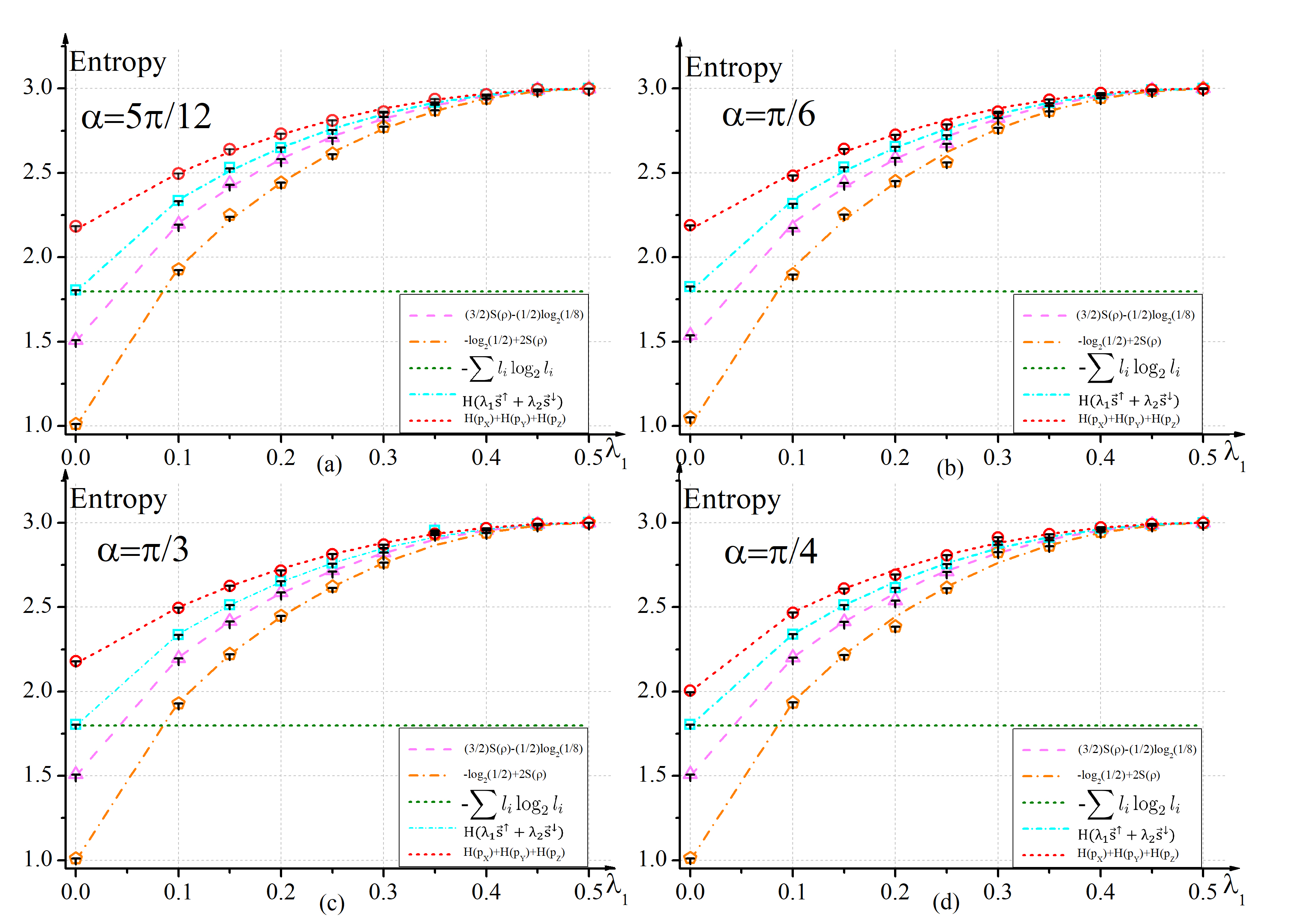}
\caption{{\bf Experimental results.} Experimental results of the measurements of three observables $\sigma_{x}$, $\sigma_{y}$ and $\sigma_{z}$ with a series of mixed states $\rho\left(\lambda_{1}, \alpha\right)$. (a) Experimental results with states $\rho(\lambda_{1},\frac{5\pi}{12})(\lambda_{1}\in\{0,0.2,0.3,0.4,0.5\})$, (b) Experimental results with states $\rho(\lambda_{1},\frac{\pi}{6})(\lambda_{1}\in\{0,0.2,0.3,0.4,0.5\})$, (c) Experimental results with states $\rho(\lambda_{1},\frac{\pi}{3})(\lambda_{1}\in\{0,0.2,0.3,0.4,0.5\})$, and (d) Experimental results with states $\rho(\lambda_{1},\frac{5\pi}{4})(\lambda_{1}\in\{0,0.2,0.3,0.4,0.5\})$. The magenta, orange, green, and blue curves correspond to the theoretical curves on the right side of the entropic uncertainty relations (\ref{13}), (\ref{14}), (\ref{15}), and (\ref{17}), respectively. The red curve represents the theoretical calculation of the left sides of four inequalities, the experimental data are represented by hollow shapes, and the short black lines are error bars with $\pm1\sigma$ standard deviation.}\label{f8}
\end{figure}

Figure \ref{f8} shows the experimental results of the measurements of three observables $\sigma_{x}$, $\sigma_{y}$, and $\sigma_{z}$ under a series of mixed states $\rho\left(\lambda_{1}, \alpha\right)$, with $ \lambda_{1}=\left\lbrace 0,0.2,0.3,0.4,0.5\right\rbrace$ and $\alpha=\left\{\frac{\pi}{6},\frac{\pi}{4}, \frac{\pi}{3}, \frac{5 \pi}{12}\right\}$. It can be seen from the figure that the experimental data points just lie on the theoretical curves, and the experimental values of the left-hand sides of the inequalities are always above the lower limits of the corresponding inequalities. The validity of all four inequalities (\ref{13}), (\ref{14}), (\ref{15}), and (\ref{17}) are verified experimentally, and the lower limit of Eq. (\ref{17}) is found to be the tightest.

\section{Conclusions}

Mixed states characterized by statistical probabilities are generally more close to the reality of the micro world. The difference of quantum behaviors between pure state and mixed state systems is normally non-trivial. In this paper, we presented an experimental test of the optimal majorization uncertainty relation for mixed qubit states, where the polarization states with varying degrees of mixedness were generated by means of two independent laser beams. The tightness of the majorization relation was verified by measuring the Lorenz curves. The experimental results show that the boundary of the majorization uncertainty relation is optimal in various polarization states of mixing degrees. Meanwhile, the experimental results show that the entropic uncertainty relations obtained from the majorization relation have more stringent lower bounds than other entropic uncertainty relations for most mixed states. Our experiment further enriches the experimental research on the uncertainty relationship of the mixed state. The method in this work provides a simple idea for preparing mixed states, and based on it simple optical elements can be used to realize the preparation of any state. The same method is also suitable for single photons. Since the mixed state in quasi-classical optics is easy to prepare, our experiment also provides a basis for further testing  of the incompatibility by generalized measurement(positive operator-valued measure, POVM), quantum intertextuality and other quantum properties in a more general (impure state) quantum system. With the rapid development of optical technology, experimental testing of non-classical effects with coherent light will certainly further deepen our understanding of the quantum uncertainty principle and its application in high-precision measurement.

Finally, it is remarkable that though the semi-classical coherent light may improve the efficiency of experiment in quantum processing, numerical simulation \cite{ZHSL} also provide an important means for relevant investigations and deserves more attention.

\section*{Acknowledgements}
\noindent
We are wholeheartedly grateful to our beloved collaborator J. L. Li for his constructive discussion on this work and others. This work was supported in part by the National Natural Science Foundation of China(NSFC) under the Grants 11375200 and 11635009; by the University of Chinese Academy of Sciences.

\end{document}